# Enzyme-free in situ polymerization of conductive polymers catalyzed by porous Au@Ag nanowires for stretchable neural electrodes

Yuyang Li[1*], Changbai Li[1*], Yangpeiqi Yi[1], Nader Marzban[1], Chengzhuo Yu[1], Tobias Abrahamsson[1], Zesheng Liu[1], Justinas Palisaitis[2], Xianjie Liu[1], Zhixing Wu[1,3,4], Eylül Ceylan[1], Per O.Å. Persson[2], Mats Fahlman[1], Xenofon Strakosas[1], Magnus Berggren,[1,5] Daniel T. Simon[1#], Klas Tybrandt[1#]

1) Laboratory of Organic Electronics, Department of Science and Technology, Linköping University, SE-60174 Norrköping, Sweden.

2) Division of Thin Film Physics, Department of Physics, Chemistry and Biology (IFM), Linköping University, Linköping SE-581 83, Sweden

3) Department of Chemical Engineering, Stanford University, Stanford, California 94305, USA

4) SUNCAT Center for Interface Science and Catalysis, SLAC National Accelerator Laboratory, Menlo Park, California 94025, USA

5) Wallenberg Initiative Materials Science for Sustainability, Department of Science and Technology, Linköping University, SE-60174 Norrköping, Sweden.

*Equal contribution.

#Corresponding authors

## Abstract

In situ polymerization of conductive polymers (CPs) represents a transformative approach in bioelectronics, by enabling the controlled growth of electrically active materials right at the tissue or device surface to create seamless biotic-abiotic interfaces. Traditional CP deposition techniques often use high anodic potentials, non-physiological electrolytes, or strong oxidants, making them harmful to adjacent tissues. A possible solution is enzymatic polymerization which operates under milder conditions, but it is limited by the stability and activity window of the enzyme catalysts, low throughput, and challenges in spatially confining polymer growth. To resolve these issues, here we developed one dimensional porous Au@Ag nanowires with horseradish peroxidase (HRP)-like catalytic properties, thereby for the first time enabling mild in situ enzyme-free polymerization of the conductive polymers near neutral pH. The enzyme-free polymerization is demonstrated both in aqueous dispersions at pH=6 and in situ onto porous Au@Ag nanowires based stretchable electrodes. Following enzyme-free catalytic polymerization, the electrically conducting polymers coating on the electrode greatly improves the impedance and achieves an impedance of 2.6 kΩ at 1 kHz for 50x50 µm large electrodes.

## Introduction

In situ polymerization of conductive polymers (CPs) has emerged as a powerful and adaptable approach for fabricating electrically active coatings directly on substrates or within composite matrices.[1] This strategy enables fine-tuned control over the electrical, mechanical, and interfacial properties of the resulting materials. Widely studied CPs such as poly(3,4-ethylenedioxythiophene):polystyrene sulfonate (PEDOT:PSS), polyaniline (PANI), polypyrrole (PPy), and poly(bis[3,4-ethylenedioxythiophene] 3-thiophene butyl sulfonate) (PETE-S) offer tunable conductivity, environmental stability, and compatibility with diverse fabrication techniques.[1,2,3] In situ synthesis is typically achieved via enzymatic polymerization,[4,5] chemical oxidative polymerization (e.g., ammonium persulfate for PANI, $Fe^{3+}$ for PPy),[6,7] electrochemical deposition,[8] or vapor-phase polymerization.[9] These methods enable precise control over polymer morphology and interfacial adhesion, making them well-suited for applications ranging from wearable electronics and energy storage to biosensors and antistatic coatings.

A key advantage of in situ polymerization is its adaptability to biological contexts. By operating under aqueous, ambient, and near-neutral pH conditions, it can preserve both cell viability and device integrity. Recent studies, such as Zhang et al.,[1] have highlighted how synthetic polymers can be directed to specific cellular domains—intracellular, membrane-bound, or extracellular—paving the way for real-time biosensing, neuromodulation, and regenerative scaffolds. For instance, hydrogel-embedded PEDOT:alginate networks formed via sulfonate doping achieve enhanced conductivity and allow for injectable biosensors.[10] Enzyme-guided localization enables selective polymer growth on targeted neuronal subtypes,[11] while photo-crosslinked conductive scaffolds help stabilize soft biointerfaces and promote synchronized tissue regeneration.[12]

Despite these advances, conventional polymerization methods pose challenges in bioelectronics. Electropolymerization often requires high voltages and harsh electrolytes and chemical oxidation involves toxic reagents and post-treatment.[13,14,15] Although enzymatic and photo-initiated polymerization routes have emerged as an attractive "green chemistry" alternative, they still facing challenges such as limited stability and lifetime of the enzymes, and that catalytic activity is highly sensitive to temperature, pH, or the presence of organic solvents or reaction byproducts.[16] Photopolymerization also includes photoinitiator molecules that can be toxic, and moreover, the physical limitation of penetration depth of often used high-energy UV light into biological tissue is severely limited to about 100 micrometers, which limits the thickness of structures that can be fabricated and makes deep-tissue in vivo polymerization practically impossible.[17] In practice, high-fidelity electrophysiological recordings still rely on noble-metal electrodes (e.g., gold, platinum),[18,19] where the interfacial quality between the conductive polymer and metal surface plays a critical role in signal transmission. Poor coupling at this interface can lead to increased impedance, signal degradation, and mechanical failure. Optimizing this interface—via surface functionalization, polymerizable tethers, or catalyst anchoring—is therefore essential for high-performance, durable devices.[20]

A potential alternative to enzymes is noble metal-based nanozymes, that artificial enzymes of nanomaterials such as gold (Au), silver (Ag), platinum (Pt), and palladium (Pd), which have gained attention as enzyme-mimicking metallic catalysts.[21,22] These nanozymes blend the robustness and surface tunability of inorganic nanoparticles with the substrate specificity and unique functionalities of natural enzymes. Of which, gold nanozymes have attracted significant attention due to their unique combination of biocompatibility, chemical stability, and surface plasmon resonance properties.[23] These features enable them to catalyze a variety of biochemical reactions, including peroxidase, oxidase, catalase, and glucose oxidase-like activities, under conditions that often denature natural enzymes. The catalytic performance of gold nanozymes can be precisely tuned by controlling particle size, shape, surface chemistry, and ligand functionalization.[24] This tunability, combined with their ease of synthesis and functionalization, has led to their integration into diverse applications. Recent advances in nanotechnology have enabled the design of hybrid and multifunctional gold nanozymes, incorporating other metals, polymers, or biomolecules to enhance catalytic efficiency and substrate specificity.[25] As research progresses, gold nanozymes are poised to become a cornerstone in next-generation catalytic systems, bridging the gap between nanomaterials science and enzyme engineering. At present, gold based nanozymes are studied mostly in the application field of biomedicine, sensing, and environmental applications.[26]

In this work, we present for the first time a near neutral-pH, enzyme-free, and electroless in situ polymerization strategy by developing horseradish peroxidase (HRP)-like porous gold coated silver nanowires (Au@Ag NWs). Leveraging the intrinsic catalytic activity of sub-20 nm gold nanoparticles arranged in a high-surface-area one dimensional network, ETE-S (4-(2-(2,5-bis(2,3-dihydrothieno[3,4-b][1,4]dioxin-5-yl)thiophen-3-yl)ethoxy)butane-1-sulfonate) monomers are polymerized under mild aqueous conditions (pH ~6, room temperature) without the need for harsh chemicals or electrical stimuli. The monomers ETE-S are polymerized onto stretchable nanowire scaffolds, forming porous, mechanically compliant electrode surfaces with significantly enhanced electroactive area. Compared to unmodified Au@Ag NWs electrodes, these hybrid constructs demonstrate over an order-of-magnitude impedance reduction and robust mechanical stability under repeated deformation, paving the way for next-generation, biocompatible, and high-performance bioelectronic systems.

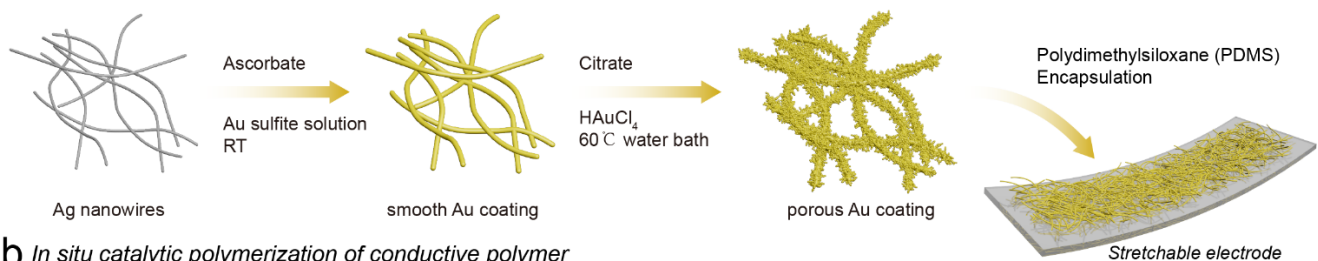

**a** *Synthesis and fabrication of Au@Ag nanowires based electrode*

**b** *In situ catalytic polymerization of conductive polymer*

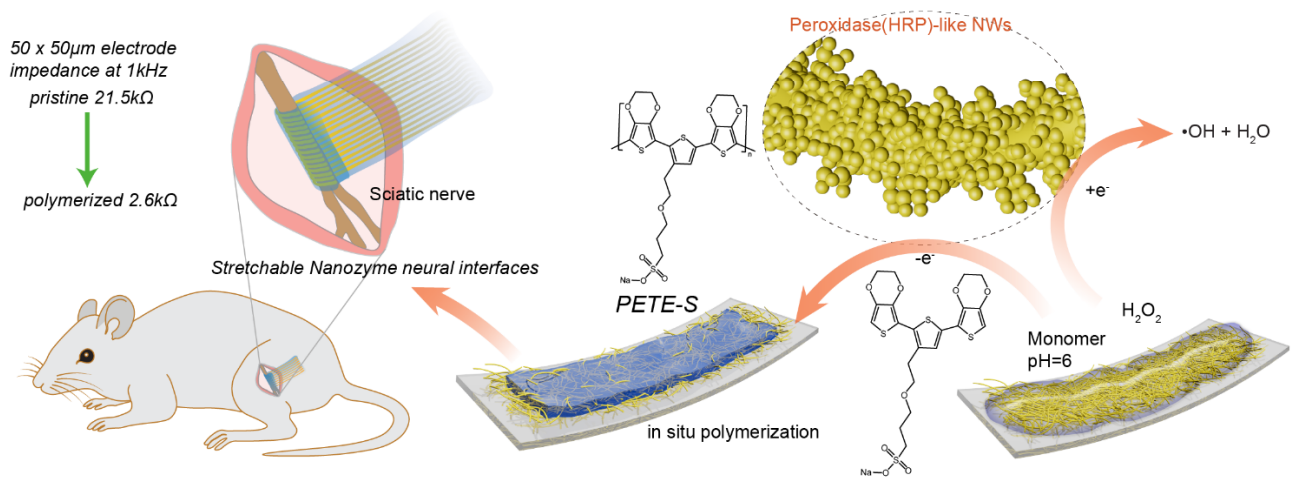

*Figure 1. Schematic illustration of synthesis procedure and application of porous Au@Ag NWs as HRP-like nanozyme. (a) Two-step Au coating on Ag nanowires template, fabricated into polydimethylsiloxane (PDMS) based stretchable electrode. (b) In situ enzyme-free HRP-like catalytic polymerization of ETE-S monomers forming conductive polymers PETE-S directly onto the stretchable electrode.*

## Results and discussion

The synthesis of porous Au@Ag NWs includes two steps. In the first step, commercial Ag nanowires (12 μm in length, 25 nm in diameter) are utilized as templates to perform first smooth Au coating layer. The method is adopted from previous report,[19],[27] typically, 30 wt% HAuCl$_4$ solution is mixed with 1 M NaOH and 0.1 M Na$_2$SO$_3$ to prepare the gold sulfite complexes in alkaline condition as the gold precursor, which has been proved to avoid galvanic replacement etching on the Ag nanowires template. The synthesis is performed at room temperature under stirring at 450 rpm, with ascorbate used as reductant. It can be seen the color of solution turns from gray to completely goldish in 20 minutes. At the second step, the smooth Au@Ag NWs synthesized from first step are being used as templates, with HAuCl$_4$ as gold precursor, sodium citrate is added as both reductant and surfactant, the reaction is performed under stirring in a 60-degree water bath for three hours. The final color turns dark gray. The morphology is depicted in the scanning electron microscope (SEM) images shown in Figure 2a-b.. As can be seen in Figure 2a, the first step synthesized smooth Au@Ag NWs are homogeneous flat surface morphology with the diameter of about 50 nm, and in Figure 2b, the second step synthesized porous Au@Ag NWs are also homogeneous in diameter around 200 nm. While a closer examination by High-Angle Annular Dark-Field Scanning Transmission Electron Microscopy (HAADF-STEM) imaging in Figure 2c shows nanowires owning the rough surface with nanoparticles coating on them. The EDX elemental mapping in Figure 2d indicates the preserved Ag core and surrounding Au nanoparticles coating, with particle sizes in sub-30 nm range, the rough surface offers high aspect surface area. The electromechanical performance of the porous Au@Ag NWs/PDMS conductors was investigated (Figures 2e and 2f). A 1 mm-width conductor was fabricated for strain-resistance testing. As shown in Figure 2e, the conductor maintains low resistance (~4 Ω) under strains up to 150%. Figure 2f demonstrates high durability, with low sheet resistance maintained over 500 cycles at 50% strain. A backlight image of the electrode under 50% strain (Figure 2h) shows a dense, continuous structure without visible cracks.

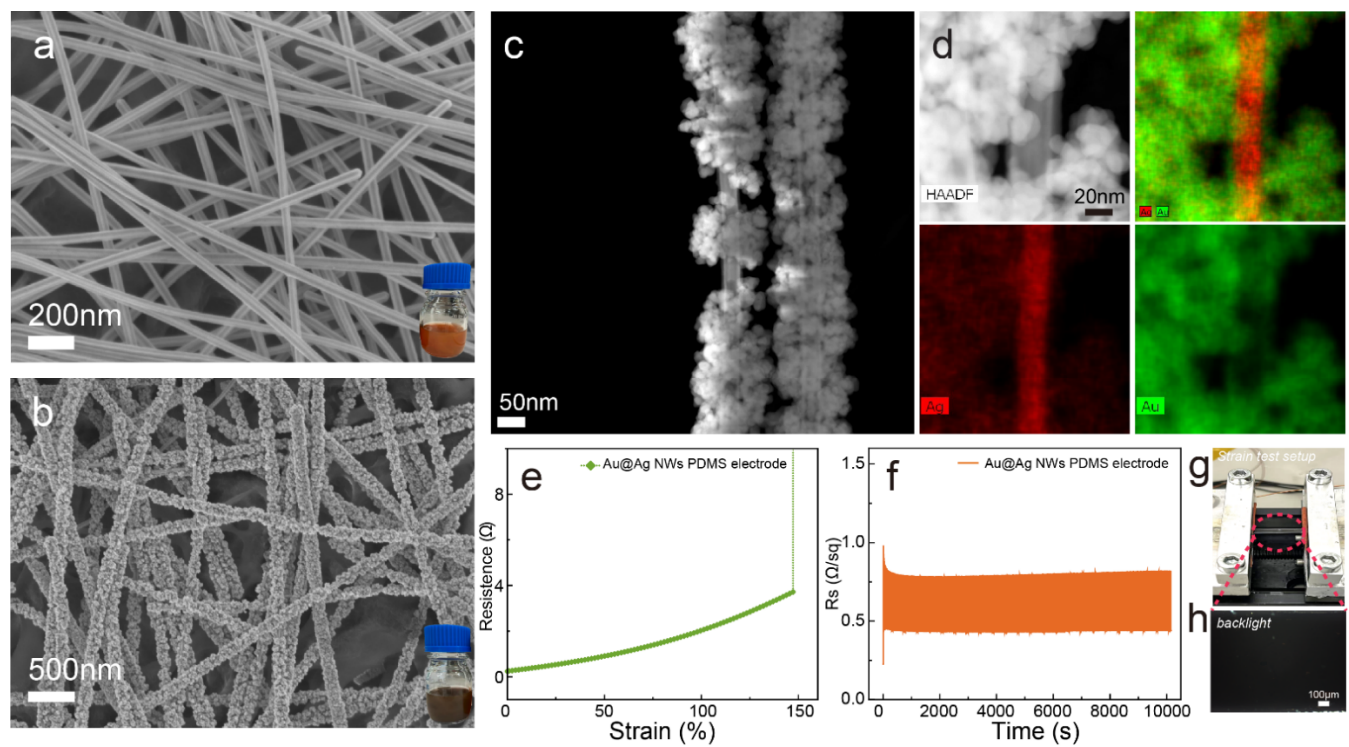

Figure 2. Morphological characterization of porous Au@Ag NWs. Scanning Electron Microscope (SEM) images of (a) first step synthesized smooth Au coated Ag nanowires, (b) second step synthesized porous Au coated Ag nanowires. (c) High-Angle Annular Dark-Field Scanning Transmission Electron Microscopy (HAADF-STEM) images of porous Au coated Ag nanowires. (d) Energy-dispersive X-ray (EDX) elemental mapping of porous Au-coated Ag nanowires revealing the distribution of Au and Ag elements. Electromechanical performance for (e) resistance-strain test and (f) sheet resistance (Rs) strain cycling under 50% strain of polydimethylsiloxane (PDMS)-porous Au@Ag NWs based stretchable electrode. (g) Strain test setup. (h) Optical backlight imaging of electrode under 50% strain.

Gold (Au) nanoparticles are well known for their enzyme-like properties, and its catalytic activity is also strongly pH dependent that can display multiple enzyme imitating including peroxidase, glucose oxidase (GOD) and catalase.[28,29,30,31,32,33] When grown on the surface of nanowires, the resulting hybrid structures combine the advantages of one-dimensional metallic conductive fillers with horseradish peroxidase (HRP)-like catalytic activity. In this study, porous Au@Ag NWs were evaluated for their HRP-like catalytic performance using the oxidation of 3,3′,5,5′-tetramethylbenzidine (TMB) by hydrogen peroxide ($H_2O_2$) as a model reaction.[34] This reaction generates an oxidized product (oxTMB) that exhibits a characteristic UV–vis absorption peak at 652 nm. As shown in Figure 3a, the catalytic activity of the porous nanowires is strongly pH-dependent, with the highest absorption intensity observed at pH 6. This suggests optimal catalytic activity under mildly acidic conditions, which is beneficial for potential applications in biological systems. In contrast, catalytic activity is significantly reduced at lower pH values. The pH dependence was also observed in the catalytic polymerization of ETE-S monomer. As shown in Figure 3b, in the absence of nanowires, $H_2O_2$ can oxidize the monomer to a limited extent at acidic pH 3, as expected. However, at pH 6, no obvious color change is observed, and UV–vis spectra confirm that the monomer peak at 350 nm remains intact across all samples. When porous Au@Ag NWs are introduced (Figure 3c), all samples exhibit a blue color, with the deepest blue appearing at pH 6. The UV–vis spectra show that at pH 6, the monomer peak disappears completely, indicating complete polymerization. At pH 4 and 3, the monomer peak intensity is reduced compared to the control (Figure 3b), confirming HRP-like catalytic activity across all tested pH values, with optimal performance at pH 6.

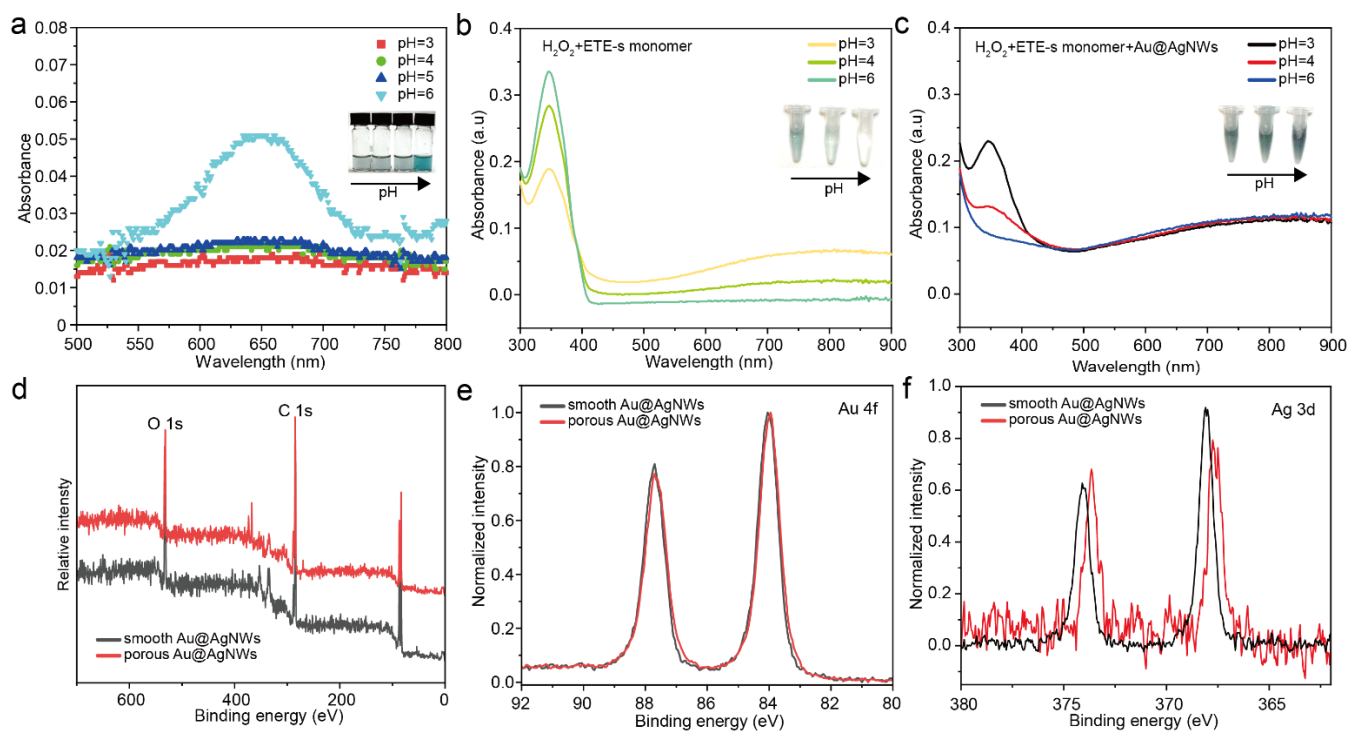

*Figure 3. Peroxidase-like catalytic performances and polymerization of ETE-S. (a) The UV–vis spectra of the peroxidase-like catalytic systems in the dependence of pH from 3 to 6, measured from the supernatant after incubation of TMB, $H_2O_2$, and porous Au@Ag NWs. (b) The UV–vis spectra of ETE-S polymerization without nanowires in the dependence of pH, measured from the supernatant after incubation of ETE-S monomer, $H_2O_2$, pH adjusted by adding HCl. (c) The UV–vis spectra of porous Au@Ag NWs catalytic ETE-S polymerization in the dependence of pH, measured from the supernatant after incubation of ETE-S monomers, $H_2O_2$, and porous Au@Ag NWs, pH adjusted by adding HCl. The inset photos in (a, b, c) show color differences. (d, e, f) X-ray Photoelectron Spectroscopy (XPS) characterization of smooth Au@Ag NWs and porous Au@Ag NWs at Au 4f and Ag 3d.*

To further investigate the catalytic properties of the porous Au@Ag NWs, X-ray photoelectron spectroscopy (XPS) measurements were conducted (Figures 3d–f). Compared to their smooth counterparts, the porous nanowires exhibit higher overall signal intensity due to increased surface area, along with stronger O 1s and C 1s peaks, indicating greater amounts of surface oxides or adsorbed species. The Au 4f spectra are similar for both samples, confirming the presence of metallic Au⁰, although the porous sample shows slightly broader peaks, suggesting subtle heterogeneity in the local chemical environment. Notably, the Ag 3d peaks in the porous sample are more intense, broadened, and shifted by approximately 0.2 eV toward lower binding energy, implying possibly electron transfer from Au to Ag and a distribution of Ag in multiple chemical states, including partially oxidized forms. Taken together, the increased surface area, altered surface chemistry, and intermetallic charge transfer in the porous Au@Ag NWs are expected to enhance their catalytic performance compared to smooth nanowires.[35]

After fabricating stretchable electrodes using porous Au@Ag NWs as the conductive filler, the in situ catalytic polymerization performance of ETE-S monomers forming conductive polymers PETE-S on the stretchable nanowire electrode was further evaluated. The fabrication method of the stretchable nanowire electrodes follows a previously reported protocol.[19] Briefly, porous Au@Ag NWs are patterned onto a PVDF membrane via vacuum filtration, achieving an area density of 2 mg/cm². A layer of PDMS is then spin-coated onto the membrane and cured on a hotplate at 70 °C for 20 minutes. After curing, the PVDF membrane is peeled off, leaving the nanowires partially embedded in the PDMS matrix. A mask is applied during the subsequent encapsulation process to define the exposed electrode area. Firstly, three parallel electrodes of 0.5 mm² exposed tip area were measured impedance as pristine state, the curves almost are overlapping as shown in Figure 4b. Then, the in situ catalytic polymerization procedure is illustrated in Figure 4a. First, a droplet of ETE-S monomer solution is casted onto the exposed electrode area, then dipping into deionized water containing 1.8 wt% $H_2O_2$, with the pH adjusted to 6–7, and incubate at room temperature for 10 minutes, subsequently the electrode is rinsed with deionized water to remove excess reactants. Electrochemical Impedance Spectroscopy (EIS) results are shown in Figure 4b. An obvious decrease in impedance is observed below 1 kHz frequency, particularly in the capacitive low-frequency region, indicating that the conductive polymer PETE-S coating enhances the capacitance of the electrode. The three electrodes are cast with different volumes of ETE-S monomers, the impedance of electrodes 2 and 3 are overlapping and lower than that of electrode 1, meaning that excess amount of monomer coating. Optical images in Figure 4e confirm the presence of a black color -polymer layer formed across the surface for half coated electrode 1 and fully coated electrode 3, compared to pristine electrode. To assess the catalytic polymerization performance at smaller electrode dimensions—relevant for neural interfacing applications—16-channel microelectrodes with individual electrode areas of 50 × 50 μm were fabricated and tested (Figure 4f). EIS measurements show a dramatic decrease in impedance at 1 kHz from pristine 21.5 kΩ to 2.6 kΩ after polymerization for 50x50 μm large electrodes as shown in Figure 4g, confirming successful conducting polymers PETE-S coating. In Figure 4f and 4i, optical images after polymerization reveal a slight darkening of the Au electrode surface, consistent with the formation of the PETE-S polymers layer. Collectively, these results highlight the excellent mechanical robustness and surface HRP-like catalytic activity of the porous Au@Ag NWs electrodes, supporting their potential for use in soft and stretchable bioelectronics. In total, EIS results provide compelling evidence for the successful catalytic polymerization of the Au electrode with conductive PETE-S polymers. Both the impedance magnitude and phase angle plots consistently show that increasing the amount of deposited conducting polymers PETE-S, that reduces the overall impedance and the charge transfer resistance, as evidenced by both the lower impedance values and the shift of the characteristic frequency to higher values, but also enhances the charge transfer kinetics at the electrode-electrolyte interface. The porous Au@Ag NWs electrode coated with conductive polymers PETE-S formed by enzyme-free polymerization demonstrates superior performance with exhibiting nearly 10 times decreased impedance at 1KHz. This enhanced electrochemical performance is highly desirable for applications such as biosensing, catalysis, and energy storage.

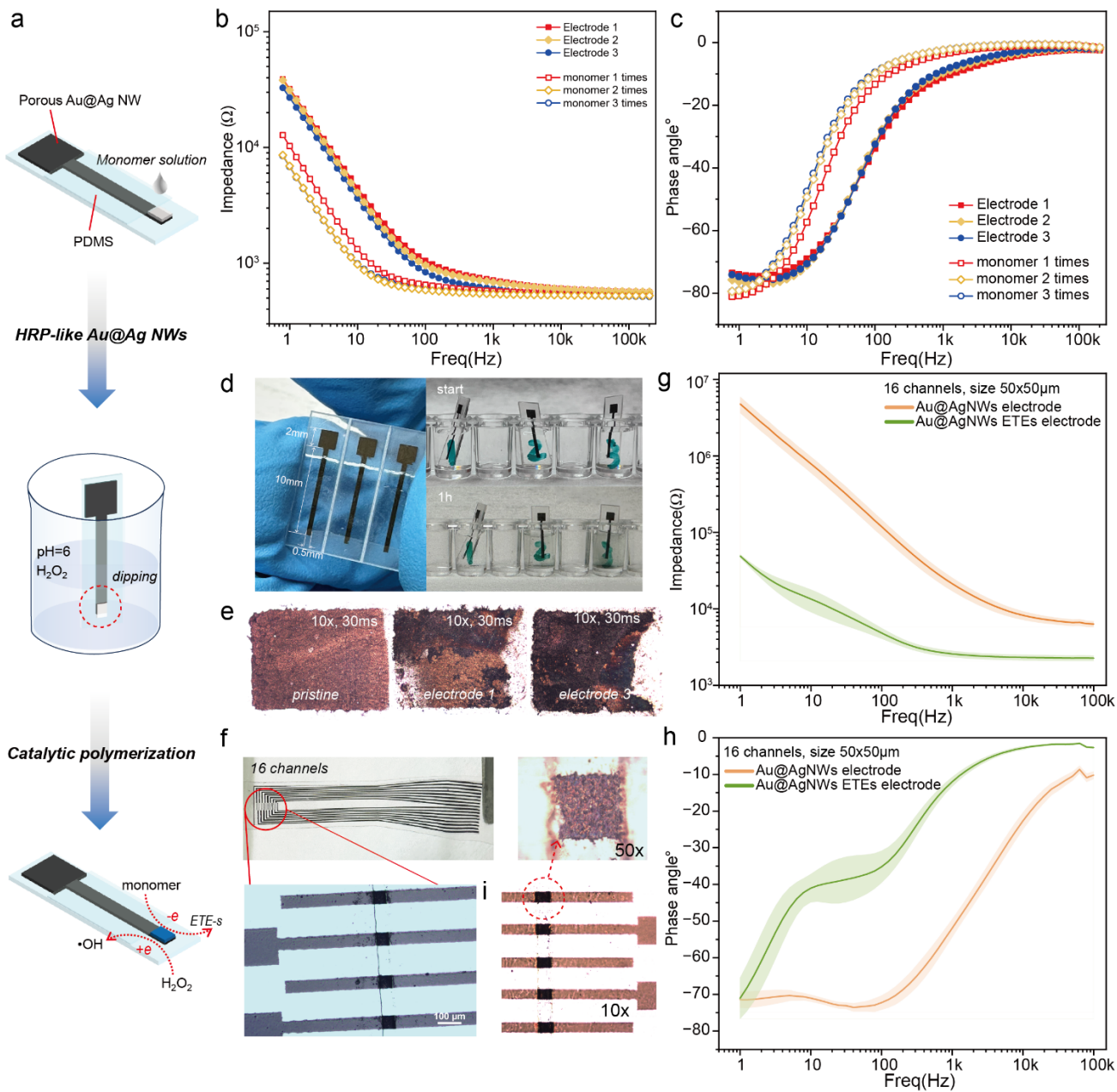

Figure 4. In situ enzyme-free catalytic polymerization of ETE-S monomer on polydimethylsiloxane (PDMS)-porous Au@Ag NWs based stretchable electrode. (a) Schematic of catalytic polymerization procedure. Electrochemical impedance spectroscopy (EIS) measures Bode plot of impedance vs. Frequency for (b) 0.5 mm width electrode before and after in situ polymerized PETE-S, and (c) phase angle vs. Frequency. (d) photos of half encapsulated Au@Ag NWs electrodes for polymerization experiments. Optical microscope images of (e) pristine Au@Ag NWs PDMS electrode, electrode 1 after polymerization with one time volume of monomer, electrode 3 after polymerization with three times volume of monomer. (f) 16 channels Au@Ag NWs PDMS electrodes with size 50x50 μm opening. Electrochemical impedance spectroscopy (EIS) measures Bode plot of impedance vs. Frequency for (b) 16 channels size 50x50 μm electrodes before and after in situ polymerized PETE-S, and (h) phase angle vs. Frequency. Optical microscope images (i) after polymerization and 50x magnification of PETE-S coated electrode.

## Conclusions

The developed catalytically active porous Au@Ag NWs, for the first time, open up new possibilities for mild in situ enzyme-free polymerization of conducting polymers both in solution and onto surfaces. The high surface porosity introduced by Au nanoparticle coating imparts pronounced HRP-like catalytic activity to the electrode. This catalytic performance is optimal at pH 6, as demonstrated by TMB oxidation assays, making it well-suited for enzyme-free polymerization reactions in biologically relevant, near-neutral environments. Importantly, in situ catalytic polymerization of conductive polymers such as PETE-S was successfully achieved on electrodes of various sizes, including small microelectrodes (50 × 50 μm), which exhibited a low impedance of 2.6 kΩ at 1 kHz, meanwhile Au@Ag NW PDMS-based electrodes demonstrated low resistance up to 150% strain. By integrating gold-nanowire-based nanozymes with neutral-condition in situ polymerization, this work establishes a seamless, low-impedance interface between conductive polymers and noble metal transducers. This strategy not only retains the excellent electrochemical performance of stretchable gold electrodes but also leverages the biocompatibility and mechanical compliance of conducting polymers, offering a promising route toward soft, chronically stable, and high-fidelity neural interfaces.

# Experimental section

**Materials synthesis:** Two-step Au coating of Ag nanowires. First, to prepare gold(I) complexing basic solution as the precursor, adapted from previous report.[19] For Au-coating the Ag nanowires, 0.396 ml of 5 mg/ml Ag nanowires were dispersed in 7.811 ml DI water, then 5.433 ml PVP 55k at 5 wt% solution was added under stirring, thereafter 0.097 ml 0.1 M $Na_2SO_3$ and 0.194 ml 1 M $C_6H_7NaO_6$ were added. Subsequently, 1.464 ml chloroauric sulfite solution was added. After stirring another 1h, the solution turns totally brick-red. The final Au-coated Ag nanowires were collected by washing with DI water three times in centrifugation and redispersed in 5 wt% PVP 55k solution. In the second step, add 2 ml smooth Au-coated Ag nanowire dispersion with 0.4 ml 50 mM sodium citrate and 1.054 ml 5 wt% PVP 55k solution, add 2.12µl $HAuCl_4$, keep stirring in a 60℃ water bath for 3h; then cool down under tap water, the final solution is completely dark gray, wash with DI water 3 times, redisperse in DI water.

**Material characterization:** Electron microscopy. Scanning Electron Microscopy (SEM) images were obtained on a Zeiss Sigma 500 equipped with the Bruker quantax EDS detector. High-angle annular dark-field Scanning Transmission Electron Microscopy (HAADF-STEM) images were recorded using the Linköping FEI Titan[3] 60-300, operated at 300 kV and equipped with a SuperX EDX spectrometer. Quantitative elemental analysis was done by EDX. UV-vis absorption was measured with Implen NanoPhotometer NP80. Electrochemical impedance spectroscopy (EIS) was measured by Gamry Interface 1010E Potentiostat.

**Measurements of catalytic activity:** UV-vis spectra and sample photo were recorded at room temperature. *TMB oxidation reaction in the dependence of pH:* Typically, 50 µL Au@AgNWs dispersion (in DI water, 0.145mg/ml), 0.9 mL DI water with 1.8 wt% $H_2O_2$ and 50 µL TMB (ELSA) were subsequently added into a vial. After an incubation of 10 min at room temperature, measure the absorbance at 652 nm (a characteristic absorbance of oxidized TMB (oxTMB)) by UV-vis spectroscopy. pH is adjusted by adding 1 M HCl.

***Catalytic polymerization of PETE-S***. *In dispersion:* 50 µL Au@AgNWs dispersion (in DI water, 0.145 mg/ml), 0.9 mL DI water with 1.8 wt% $H_2O_2$ and 0.06 wt‰ ETE-S monomers were subsequently added into a vial. After an incubation of 10 minutes at room temperature, measure by UV-vis spectroscopy. pH is adjusted by adding 1 M HCl. For reference samples, replace Au@AgNWs dispersion by adding 50 µL DI water, pH is adjusted by adding 1 M HCl. *On electrode*: typically, drop casting 10 µl 0.01 wt% ETE-S monomer on to exposed electrode area, dipped in 1 ml 1.8 wt% $H_2O_2$, incubate for 10 minutes up to 1h at room temperature. Then rinse the electrode with DI water, dried at ambient condition for further characterization.

# Acknowledgements

This work was financially supported by the European Research Council (CoG 2021 Klas Tybrandt, 101089075), the Swedish Foundation for Strategic Research (FFL18-0206), the Swedish Research Council (2023-04694), the Knut and Alice Wallenberg Foundation, the Swedish Government Strategic Research Area in Materials Science on Advanced Functional Materials at Linköping University (Faculty Grant SFO-Mat-LiU No. 2009-00971). The Swedish Research Council and the Swedish Foundation for

Strategic Research are acknowledged for access to ARTEMI, the Swedish National Infrastructure in Advanced Electron Microscopy (2021-00171 and RIF21-0026).

## Author contributions

## References

1. Zhang, A., Zhao, S., Tyson, J., Deisseroth, K. & Bao, Z. Applications of synthetic polymers directed toward living cells. *Nat. Synth.* **3**, 943–957 (2024).

2. Li, C. *et al.* Engineering Conductive Hydrogels with Tissue-like Properties: A 3D Bioprinting and Enzymatic Polymerization Approach. *Small Sci.* **4**, 2400290 (2024).

3. Stavrinidou, E. *et al.* In vivo polymerization and manufacturing of wires and supercapacitors in plants. *Proc. Natl. Acad. Sci.* **114**, 2807–2812 (2017).

4. Strakosas, X. *et al.* Metabolite-induced in vivo fabrication of substrate-free organic bioelectronics. *Science (80-. ).* **379**, 795–802 (2023).

5. Kadokawa, J. ichi & Kobayashi, S. Polymer synthesis by enzymatic catalysis. *Current Opinion in Chemical Biology* vol. 14 145–153 at https://doi.org/10.1016/j.cbpa.2009.11.020 (2010).

6. Ismail, R. *et al.* Acid-assisted polymerization: the novel synthetic route of sensing layers based on PANI films and chelating agents protected by non-biofouling layer for Fe2+ or Fe3+ potentiometric detection. *J. Mater. Chem. B* **11**, 1545–1556 (2023).

7. Imani, A., Farzi, G. & Ltaief, A. Facile synthesis and characterization of polypyrrole-multiwalled carbon nanotubes by in situ oxidative polymerization. *Int. Nano Lett.* **3**, 52 (2013).

8. Roselló-Márquez, G., García-García, D. M., Cifre-Herrando, M. & García-Antón, J. Electropolymerization of PPy, PEDOT, and PANi on WO₃ nanostructures for high-performance anodes in Li-ion batteries. *Heliyon* **10**, (2024).

9. Evans, D. *et al.* Structure-directed growth of high conductivity PEDOT from liquid-like oxidant layers during vacuum vapor phase polymerization. *J. Mater. Chem.* **22**, 14889–14895 (2012).

10. Montazerian, H. *et al.* Boosting hydrogel conductivity via water-dispersible conducting polymers for

injectable bioelectronics. *Nat. Commun.* **16**, 3755 (2025).

11.  Liu, J. *et al.* Genetically targeted chemical assembly of functional materials in living cells, tissues, and animals. *Science (80-. ).* **367**, 1372–1376 (2020).

12.  Mehrotra, S., Dey, S., Sachdeva, K., Mohanty, S. & Mandal, B. B. Recent advances in tailoring stimuli-responsive hybrid scaffolds for cardiac tissue engineering and allied applications. *J. Mater. Chem. B* **11**, 10297–10331 (2023).

13.  Poverenov, E., Li, M., Bitler, A. & Bendikov, M. Major Effect of Electropolymerization Solvent on Morphology and Electrochromic Properties of PEDOT Films. *Chem. Mater.* **22**, 4019–4025 (2010).

14.  Tsong, T. Y. Electroporation of cell membranes. *Biophys. J.* **60**, 297–306 (1991).

15.  Holze, R. Overoxidation of Intrinsically Conducting Polymers. *Polymers* vol. 14 at https://doi.org/10.3390/polym14081584 (2022).

16.  Ramanavicius, S. & Ramanavicius, A. Charge Transfer and Biocompatibility Aspects in Conducting Polymer-Based Enzymatic Biosensors and Biofuel Cells. *Nanomaterials* vol. 11 at https://doi.org/10.3390/nano11020371 (2021).

17.  Bagheri, A. & Jin, J. Photopolymerization in 3D Printing. *ACS Appl. Polym. Mater.* **1**, 593–611 (2019).

18.  Tybrandt, K. *et al.* High-Density Stretchable Electrode Grids for Chronic Neural Recording. *Adv. Mater.* **30**, e1706520 (2018).

19.  Seufert, L. *et al.* Stretchable Tissue-Like Gold Nanowire Composites with Long-Term Stability for Neural Interfaces. *Small* **20**, 2402214 (2024).

20.  Boufidis, D., Garg, R., Angelopoulos, E., Cullen, D. K. & Vitale, F. Bio-inspired electronics: Soft, biohybrid, and "living" neural interfaces. *Nat. Commun.* **16**, 1861 (2025).

21.  Chen, J. *et al.* Glucose-oxidase like catalytic mechanism of noble metal nanozymes. *Nat. Commun.* **12**, 1–9 (2021).

22.  Liang, M. & Yan, X. Nanozymes: From New Concepts, Mechanisms, and Standards to Applications. *Acc. Chem. Res.* **52**, 2190–2200 (2019).

23.  Lou-Franco, J., Das, B., Elliott, C. & Cao, C. Gold Nanozymes: From Concept to Biomedical Applications. *Nano-Micro Lett.* **13**, 10 (2020).

24.  Huang, S., Xiang, H., Lv, J., Guo, Y. & Xu, L. Propelling gold nanozymes: catalytic activity and biosensing applications. *Anal. Bioanal. Chem.* **416**, 5915–5932 (2024).

25.  Tong, P.-H., Wang, J.-J., Hu, X.-L., James, T. D. & He, X.-P. Metal–organic framework (MOF) hybridized gold nanoparticles as a bifunctional nanozyme for glucose sensing. *Chem. Sci.* **14**, 7762–7769 (2023).

26.  Huang, Y., Ren, J. & Qu, X. Nanozymes: Classification, Catalytic Mechanisms, Activity Regulation, and Applications. *Chem. Rev.* **119**, 4357–4412 (2019).

27.  Hong, X., Tan, C., Chen, J., Xu, Z. & Zhang, H. Synthesis, properties and applications of one- and two-dimensional gold nanostructures. *Nano Research* vol. 8 40–55 at https://doi.org/10.1007/s12274-014-0636-3 (2015).

28. Ball, L. T., Lloyd-Jones, G. C. & Russell, C. A. Gold-catalyzed direct arylation. *Science (80-. ).* **337**, 1644–1648 (2012).

29. Zeng, D. *et al.* Gold nanoparticles-based nanoconjugates for enhanced enzyme cascade and glucose sensing. *Analyst* **137**, 4435–4439 (2012).

30. Drozd, M., Pietrzak, M., Parzuchowski, P. G. & Malinowska, E. Pitfalls and capabilities of various hydrogen donors in evaluation of peroxidase-like activity of gold nanoparticles. *Anal. Bioanal. Chem.* **408**, 8505–8513 (2016).

31. Corma, A. & Garcia, H. Supported gold nanoparticles as catalysts for organic reactions. *Chem. Soc. Rev.* **37**, 2096–2126 (2008).

32. Zhang, R., Yan, X., Gao, L. & Fan, K. Nanozymes expanding the boundaries of biocatalysis. *Nat. Commun.* **16**, 6817 (2025).

33. Chen, J. *et al.* Glucose-oxidase like catalytic mechanism of noble metal nanozymes. *Nat. Commun.* **12**, 3375 (2021).

34. Wang, C. *et al.* Structural Regulation of Au-Pt Bimetallic Aerogels for Catalyzing the Glucose Cascade Reaction. *Adv. Mater.* 2405200 (2024) doi:10.1002/adma.202405200.

35. Liu, J.-H., Wang, A.-Q., Chi, Y.-S., Lin, H.-P. & Mou, C.-Y. Synergistic Effect in an Au−Ag Alloy Nanocatalyst: CO Oxidation. *J. Phys. Chem. B* **109**, 40–43 (2005).